\documentclass[prb,twocolumn,showpacs,preprintnumbers,amsmath,amssymb,showkeys,citeautoscript,floatfix]{revtex4-1}

\usepackage{graphicx, bm, hyperref,color}

\renewcommand{\vec}[1]{\boldsymbol #1}
\newcommand{\e}{\text{e}}

\def\12{\frac{1}{2}}

\newcommand{\be}{\begin{equation}}
\newcommand{\ee}{\end{equation}}
\newcommand{\bea}{\begin{eqnarray}}
\newcommand{\eea}{\end{eqnarray}}

\renewcommand\Im{\operatorname{Im}}

\begin{document}

\title{Study of the cavity-magnon-polariton transmission line shape}

\author{M. Harder\footnote{Electronic address: michael.harder@umanitoba.ca}, Lihui Bai, C. Match, J. Sirker and C.-M. Hu}

\affiliation{Department of Physics and Astronomy, University
of Manitoba, Winnipeg, Canada R3T 2N2}

\date{\today}

\begin{abstract}

We experimentally and theoretically investigate the microwave transmission line shape of the cavity-magnon-polariton (CMP) created by inserting a low damping magnetic insulator into a high quality 3D microwave cavity.  While fixed field measurements are found to have the expected Lorentzian characteristic, at fixed frequencies the field swept line shape is in general asymmetric.  Such fixed frequency measurements demonstrate that microwave transmission can be used to access magnetic characteristics of the CMP, such as the field line width $\Delta H$.  By developing an effective oscillator model of the microwave transmission we show that these line shape features are general characteristics of harmonic coupling.  At the same time, at the classical level the underlying physical mechanism of the CMP is electrodynamic phase correlation and a second model based on this principle also accurately reproduces the experimental line shape features.  In order to understand the microscopic origin of the effective coupled oscillator model and to allow for future studies of CMP phenomena to extend into the quantum regime, we develop a third, microscopic description, based on a Green's function formalism.  Using this method we calculate the transmission spectra and find good agreement with the experimental results.

\end{abstract}


\maketitle
\section{Introduction}
Strong light-matter interactions in condensed matter systems are a rich source of physics, underlying such important concepts as the polariton \cite{Mills1974} while holding the key to new technological development, such as quantum information processing.  In this direction much work has recently been devoted to the strong magnon-photon interactions between low loss magnetic materials and high quality microwave cavities/resonators \cite{Huebl2013a, Tabuchi2014a, Zhang2014, Bai2015, Cao2014, Goryachev2014, Tabuchi2015, Tabuchi2015b, Haigh2015a, Rameshti2015, Bai2015b, Lambert2015, Lambert2015b, Hu2015, Osada2015, Haigh2015b, Zhang2015b, Bourhill2015, Yao2015, Zhang2015g, MaierFlaig2016},  motivated by the potential for large coherent coupling in magnetically ordered systems \cite{Soykal2010, Soykal2010a}.  Initial experiments demonstrated such phenomena at low temperatures \cite{Huebl2013a, Tabuchi2014a} however strong spin-photon coupling at room temperature was soon realized \cite{Zhang2014, Bai2015}.  Such work has provided the foundation for many exciting new possibilities such as cavity mediated coupling of spatially separated magnetic moments \cite{Lambert2015b}, cavity mediated qubit-magnon coupling \cite{Tabuchi2015, Tabuchi2015b}, the merging of quantum optics and spintronics through the use of whispering gallery modes \cite{Osada2015, Haigh2015b, Zhang2015b, Bourhill2015}, the possible integration of microwave, optical and magnonic systems \cite{Zhang2015f} and the development of magnon dark mode memory architectures \cite{Zhang2015g}.

While much of the recent work has focused on the potential of strong spin-photon coupling for quantum information technologies, from the spintronics perspective \cite{Hu2015} these interactions are also remarkably interesting since they may be detected both optically through microwave transmission and electrically via spin pumping measurements \cite{Bai2015, MaierFlaig2016}.  This is made even more important due to the coherent nature of the spin-photon coupling via the cavity-magnon-polariton (CMP) which originates in the electrodynamic coupling between ferromagnetic resonance (FMR) and the microwave cavity mode\cite{Bai2015}.  In spintronic systems coherence phenomena have previously been manifested in various line shape symmetries\cite{Wirthmann2010, Harder2011a, Azevedo2011} and therefore as a further step towards coherent spin current control it is important to develop a better understanding of the CMP line shape.  A detailed experimental and theoretical analysis of the CMP microwave transmission line shape is the focus of this work.  

Previously the microwave transmission spectra in strongly coupled spin-photon systems has been described using the input-output formalism of quantum optics \cite{Huebl2013a}, where the input and output microwave powers are determined by the photon occupation number \cite{WallsBook, Clerk2010}.  However typical spintronic systems operate in the semi classical regime, and although the input-output formalism becomes semi classical in the absence of anharmonicity, an explicitly classical description based on simple physical considerations which is easily extendable to multi mode coupling would provide a practical and useful analysis tool for spintronic applications.  In this context we present such a description which accurately explains our experimental findings.  Nevertheless the link between the classical and quantum descriptions is of interest and can be systematically studied by applying a Green's function formalism to the microscopic theory.  Within this framework we again accurately reproduce our experimental results.

To properly characterize a strongly coupled magnon-photon system it is necessary to measure both the field and frequency dependence of the microwave transmission, performing a two dimensional measurement $|S_{21}\left(\omega, H\right)|^2$.  In Sec. \ref{sec:exp} we describe the experimental details of such measurements and highlight the important line shape features, examining both the fixed field, $|S_{21}\left(\omega\right)|^2$, and fixed frequency, $|S_{21}\left(H\right)|^2$, line shapes.  In particular we find that while $|S_{21}\left(\omega\right)|^2$ exhibits two symmetric resonances, corresponding to the anti crossing of the coupled cavity/FMR mode, the $|S_{21}\left(H\right)|^2$ line shape is generally asymmetric and this asymmetry is controllable by the microwave frequency $\omega$.  To explain these line shape features in a general context, in Sec. \ref{sec:osc} we present a coupled oscillator model which accurately captures the key experimental features and can easily be generalized to describe multiple cavity/multiple spin wave mode coupling with several cavity driving frequencies.  While the electrodynamic origin of the CMP is known from our previous work \cite{Bai2015}, only the dispersion and line width, and not the actual microwave transmission spectra, have been studied in this context.  In Sec. \ref{sec:phase} we further apply our dynamic phase correlation model to describe the spectral line shape and accurately reproduce the experimental features, explaining the physical origin of the CMP line shape.  In Sec. \ref{sec:greens} we develop a microscopic quantum theory of the coupled spin-photon system and discuss the approximations which lead to an effective coupled oscillator model as studied in the previous sections. To make our models easier to use from a practical point of view, in Sec. \ref{sec:simp} we look at the limits of small damping and derive analytic expressions for $|S_{21}\left(\omega\right)|^2$ and $|S_{21}\left(H\right)|^2$ showing that $|S_{21}\left(\omega\right)|^2$ is well described by a Lorentzian while $|S_{21}\left(H\right)|^2$ is explicitly asymmetric.  Finally in Sec. \ref{sec:dis} we show the excellent quantitative agreement between all three models and the dispersion and line width of the CMP.

\section{Transmission spectra of the Cavity-Magnon-Polariton}\label{sec:exp}

\begin{figure}[!h]
\centering
\includegraphics[width=8.5cm]{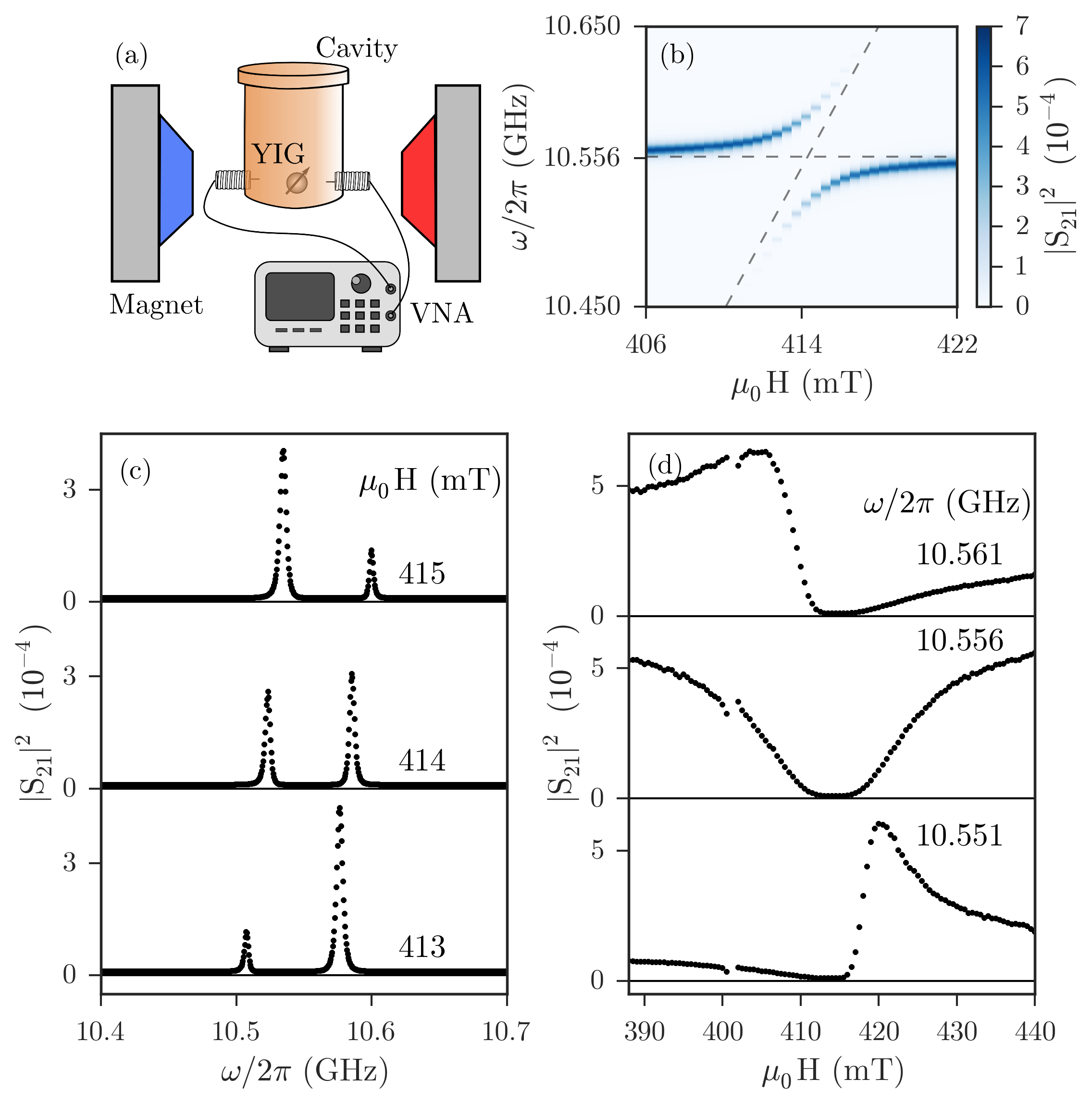}
\caption{Transmission of a strongly coupled cavity-magnon-polariton (CMP) system. (a) Experimental setup:  a YIG sphere inside a microwave cavity is placed inside a static magnet field and the microwave transmission is measured using a VNA.  (b) The full $\omega-H$ dispersion of the CMP system displays a large anti crossing characteristic of the strong coupling regime.  Horizontal and diagonal dashed lines show the uncoupled cavity and FMR modes respectively.  (c) Fixed field and (d) frequency cuts made above, at and below the coupling point, $\omega_r = \omega_c$.}
\label{dataFig}
\end{figure}

To study the CMP line shape we placed a 1 mm diameter YIG sphere (Ferrisphere Inc.) inside of an in-house cylindrical microwave cavity (2.5 cm $\times$ 2.9 cm, diameter $\times$ height) made from oxygen free copper.  As shown in Fig.~\ref{dataFig} (a) the cavity and YIG sample are placed inside of a static magnetic field $H$, which acts as a bias for the FMR, and the microwave transmission is then measured using a vector network analyzer (VNA).  Our cavity has been designed to have a TM$_{011}$ mode at a loaded resonance frequency of $\omega_c/2\pi = 10.556$ GHz (0.2\% redshifted from the unloaded value).  The intrinsic loss rate of this mode, $\beta =  \Delta \omega/\omega_c = 3 \times 10^{-4}$, is determined from its half-width-half-maximum (HWHM), $\Delta \omega$, far from coupling and corresponds to a loaded cavity quality of $Q = 1/2\beta = 1700$.  The TM$_{011}$ mode is well separated from other TE or TM modes (by more than 1.5 GHz) and 
has a circular field profile, allowing the coupling to be maximized when the microwave and static fields are perpendicular \cite{Bai2015b}. 

The YIG FMR can be tuned by the $H$ field and follows a linear dispersion, $\omega = \gamma \left(\text{H} + \text{H}_A\right)$, where $\gamma = 2 \pi \times 28 ~\mu_0$ GHz/T is the gyromagnetic ratio and  $\mu_0 \text{H}_A = -37.7$ mT is the geometry specific anisotropy field.  By tuning the FMR frequency to $\omega_c$ we observe the CMP as shown by the avoided crossing in Fig.~\ref{dataFig} (b).  Here the horizontal and diagonal dashed lines are the uncoupled cavity and FMR dispersions respectively.  Due to the high Q, small Gilbert damping ($\alpha = 0.8 \times 10^{-4}$) and large number of spins ($N_s \approx 5.2 \times 10^{18}$) the data in Fig.~\ref{dataFig} (b) showing $|S_{21}\left(\omega, H\right)|^2$ is clearly in the strongly coupled regime and we observe a dispersion gap between the upper and lower branches of $\omega_\text{gap}/2\pi = 63$ MHz.  

In Fig.~\ref{dataFig} (c) we focus on line cuts made at fixed $H$, $|S_{21}\left(\omega\right)|^2$.  These fixed field line cuts show both branches of the dispersion and the anti crossing behaviour can be seen by the $\omega_\text{gap}$ peak separation at the coupling point $\omega_r \left( \mu_0 H_{rc}\right) = \omega_r\left(414~ \text{mT} \right) = \omega_c$ where $H_{rc}$ is defined as the FMR field when $\omega_r = \omega_c$.  Above (below) the coupling point the amplitude of the upper (lower) branch decreases sharply as the cavity mode moves away from the FMR frequency and can no longer effectively drive precession.  Both branches have a clear Lorentz line shape as one would expect for such a resonant process and are observed to satisfy $|S_{21}\left(\omega_r - \delta_\omega, H_{rc} + \delta_H\right)|^2 = |S_{21}\left(\omega_r + \delta_\omega, H_\text{rc} - \delta_H\right)|^2$ for the field and frequency detunings, $\delta_H$ and $\delta_\omega$ respectively. 

Turning to the fixed frequency cuts highlighted in Fig.~\ref{dataFig} (d), $|S_{21}\left(H\right)|^2$, we see a distinctly different line shape compared to Fig.~\ref{dataFig} (c).  First, at $\omega = \omega_c$ we have a broad symmetric dip rather than a peak.  However the more striking distinction occurs when $\omega \ne \omega_c$ and the transmission line shape $|S_{21}\left(H\right)|^2$ becomes asymmetric.  This symmetry change occurs immediately away from the cavity frequency, as indicated by the upper and lower curves in Fig.~\ref{dataFig} (d) which are taken just 0.05 \% away from $\omega_c$.  We can also see that the polarity of this asymmetry changes above and below $\omega_c$, satisfying $|S_{21}\left(\omega_c - \delta_\omega, H_{rc} + \delta_H\right)|^2 = |S_{21}\left(\omega_c + \delta_\omega, H_{rc} - \delta_H \right)|^2$.  The key difference between this expression and the analogue for fixed fields is that $\omega_c$ is field independent, whereas $\omega_r$ depends on the field.  We also should note that the additional feature near 404 mT which does not obey this relationship, due to the coupling between the cavity and a spin wave mode, has been studied in detail elsewhere \cite{MaierFlaig2016} and is not of immediate interest for our line shape discussion.  

The asymmetry observed at fixed frequency is of interest for two key reasons.  First, at fixed frequency $|S_{21}\left(H\right)|$ can be used to study the magnetic characteristics of the CMP, such as the resonance field $H_r$ and field line width $\Delta H$.  Previously this information has only been examined through spin pumping\cite{Bai2015}.  In order to extract this information the transmission spectra must be fit and it is clear from Fig.~\ref{dataFig} (d) that this cannot be done using a Lorentz function as is usual for transmission measurements.  Therefore from a practical viewpoint it is important to determine the exact line shape that should be used.  Second, studies of spin rectification and spin pumping in spintronic systems have previously shown that line shape symmetries and in particular changes in line shape symmetry reveal relevant phase information\cite{Harder2011a}, in those cases regarding the relative phase between rf electric and magnetic fields.  We speculate that understanding line shape symmetries in the CMP system may provide new insight into CMP phase coherence, which would be relevant for the application of strong magnon-photon coupling to the coherent control of spin currents.  The general description of line shape symmetries presented here is the first step in this direction.

To summarize the key line shape features, we observe from (c) and (d) that i) fixed $H$ cuts are symmetric in $\omega$ ii) the amplitude of the upper (lower) branch resonance decreases rapidly above (below) the coupling point iii) the fixed $\omega$ cuts have a symmetric dip at $\omega_c$ but otherwise are generally asymmetric in $H$ and iv) the polarity of the asymmetry in the fixed $\omega$ cuts changes above and below $\omega_c$.  These line shape characteristics are universal, having been observed in many different FM samples and cavity geometries and are reproduced by all three models we now consider.  
\section{Overview and Highlights of Three Cavity-Magnon-Polariton Models}\label{sec:key}
Before describing the details of our three models, we briefly summarize the key motivations for and highlights of each approach. 

\textit{Model I: Harmonic Coupling} --  In the absence of anharmonicity the magnon-photon system is semiclassical and therefore the key features of the CMP should be contained within a coupled oscillator model.  Though this may be anticipated, such an approach has never been compared directly to the transmission spectra line shape.  While the eigenmodes of the coupled oscillator are sufficient to describe the CMP dispersion and line width evolution, in order to model the transmission spectra coupling to the input and output ports of the VNA must be included.  We perform such modelling and find that the transmission spectra of the CMP, including line shape, dispersion and damping evolution is accurately described.  One advantage of this approach is a direct demonstration of the harmonic nature of recent CMP observations.  A second advantage is the ability to easily extend the model to describe multiple cavity or magnetic modes which has recently become a subject of interest \cite{Zhang2015g, MaierFlaig2016}.  

\textit{Model II: Dynamic Phase Correlation} -- A limitation of the harmonic oscillator model is that it does not explain the physical origin of the CMP coupling.  Our previous work has shown that the CMP results from the dynamic phase correlation between Amp\`ere's and Faraday's Laws \cite{Bai2015} however only the dispersion and line width evolution has been studied in this context.  In our second model we extend this previous work and describe the transmission spectra based on the \textit{physical principle} of dynamic phase correlation using the \textit{technical tool} of microwave circuit theory.  In this approach the physical origin of coupling is made clear, which can again be easily extended to multi mode systems.

\textit{Model III: Microscopic Theory} -- The harmonic coupling and dynamic phase correlation models are both based on a macroscopic, classical starting point.  Therefore, although these models accurately describe current CMP observations and can even be applied to multi mode systems, the extension to a microscopic, quantum description is not immediately available. To provide such an approach we examine the transmission spectra based on the microscopic quantum model for the coupled spin and photon degrees of freedom \cite{Soykal2010, Soykal2010a, Huebl2013a}. Using linear spin-wave theory we show that the microscopic model reduces to a system of coupled oscillator modes. The transmission line shape can then be obtained from the full spectral function of the cavity photons. An advantage of this approach is that it can be extended step by step in order to systematically study novel effects, such as radiative damping.  

To summarize: All three models accurately describe the microwave transmission spectra, providing an important tool for the analysis of strongly coupled spin photon systems.  These models also provide an important starting point for future work -- the harmonic coupling and dynamic phase correlation models can be extended in a straightforward way to analyze multi mode systems which are currently becoming of interest, while the microscopic model can be systematically extended to examine higher order quantum effects.  
\section{Model I: Harmonic Coupling}\label{sec:osc}

\begin{figure}[h]
\centering
\includegraphics[width=8.5cm]{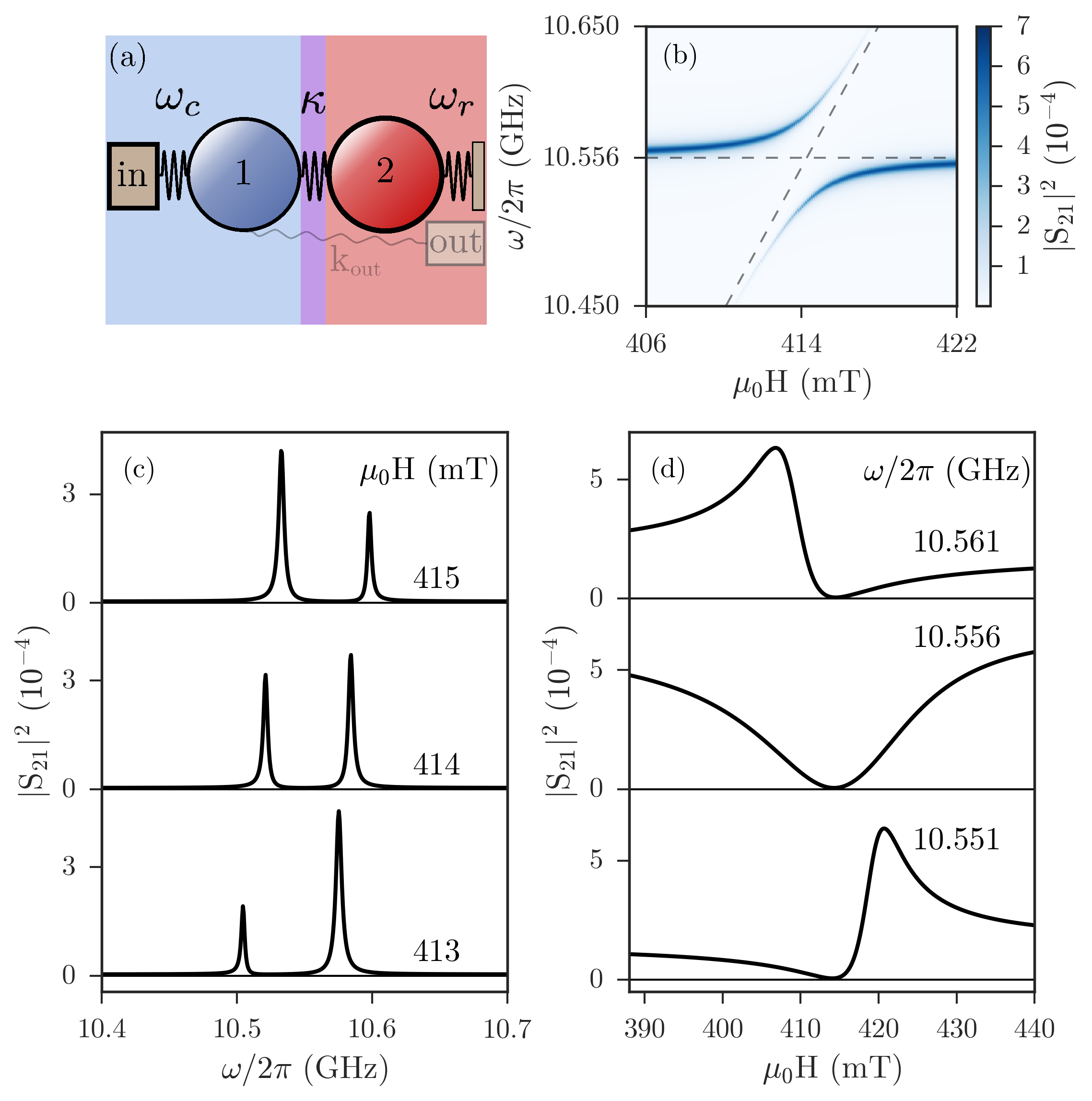}
\caption{Results of the harmonic coupling model. (a) Schematic illustration of the coupled oscillators.  The cavity is represented by the blue oscillator 1 while the FMR is represented by the red oscillator 2.  The purple region in-between indicates the coupling between the two systems. (b) The full $\omega-H$ dispersion calculated according to Eq.~\eqref{eq:oscT} using the experimental parameters for $\alpha, \beta, \omega_c$ and $\omega_r$ with $\kappa$ measured from the experimental dispersion gap and the amplitude determined from fitting.  (c) Fixed field and (d) frequency cuts made above, at and below the coupling point $\omega_r = \omega_c$ calculated according to Eq.~\eqref{eq:oscT}.}
\label{oscFig}
\end{figure}

To model the CMP we consider the coupled oscillators shown in Fig.~\ref{oscFig} (a).  Here two equal mass $(m)$ oscillators, labelled by 1 and 2, are coupled together via a spring $\kappa$.  Oscillator 1 represents the cavity and is connected to an input plunger via a spring with resonance frequency $\omega_c$.  The plunger is driven in constant motion so that $x_\text{in} (t) = x_\text{in} e^{-i\omega t}$ which produces a driving force on oscillator 1, $f(t) = \omega_c^2 x_\text{in}(t)$.  This is analogous to the constant microwave input to the cavity by the VNA.  Oscillator 2 represents the FMR and is attached to a fixed wall with a spring of resonant frequency $\omega_r$.  We add a damping coefficient of $\beta$ to oscillator 1 and $\alpha$ to oscillator 2 to model the intrinsic losses of the cavity and the losses due to Gilbert damping respectively.  There is no damping associated with the input.  

In order to extract the CMP dispersion and line width evolution it is sufficient to set $x_\text{in}  = 0$ and solve the corresponding eigenvalue problem; that is, the dispersion and line width are determined solely by the coupling $\kappa$ and the uncoupled resonances and damping, $\omega_c, \omega_r, \beta$ and $\alpha$.  The energy absorption of the coupled oscillators can also be calculated in a standard way using the dissipative function of the system \cite{LandauBookMechanics}.  However to determine the line shape of the \textit{transmission} spectra we need to calculate the input and output energy of the system.  This requires the addition of an output which models the second port of the VNA which we achieve by connecting an absorber to oscillator 1 using a spring $k_\text{out}$ as shown in Fig.~\ref{oscFig} (a).  The transmission through the system is then $|S_{21}|^2 = E_\text{out}/E_\text{in}$ where $E_\text{out}$ and $E_\text{in}$ are the kinetic energies of the output and input oscillators respectively.  Due to the high quality of our cavity, the output coupling $k_\text{out} \ll 1$, and in particular $k_\text{out} \ll \kappa$.  There is also no damping associated with the output.  

The equations of motion for oscillator 1, 2 and the output are respectively
\begin{subequations}
\begin{gather}
\ddot{x}_1 + \omega_c^2 x_1 + 2 \beta \omega_c \dot{x}_1 - \kappa^2 \omega_c^2 x_2 = f e^{-i\omega t}, \\
\ddot{x}_2 +\omega_r^2 x_2 + 2 \alpha \omega_c \dot{x}_2 - \kappa^2 \omega_c^2 x_1 = 0, \\
\ddot{x}_\text{out} - k_\text{out}^2 \omega_c^2 x_1 = 0.
\end{gather}
\label{eq:initialHarmonic}%
\end{subequations}
Here we have defined the damping coefficients $\alpha$ and $\beta$ as well as the couplings $\kappa$ and $k_\text{out}$ to be dimensionless by normalizing to $\omega_c$.  Additionally since $k_\text{out} \ll 1$ the action of the output on oscillator 1 does not need to be included in Eq.~\eqref{eq:initialHarmonic} (a).  Taking $\left(x_1, x_2, x_\text{out}\right) = \left(A_1, A_2, A_\text{out}\right) e^{-i\omega t}$, Eqs.~\eqref{eq:initialHarmonic} (a) and (b) can be written in the matrix form $\pmb{\Omega} \textbf{A} = \textbf{f}$ where $\textbf{A} = \left(A_1, A_2\right)$, $\textbf{f} = \left(-f, 0\right)$ and 

\begin{equation*}
\pmb{\Omega} = \left(\begin{array}{cc}
\omega^2 - \omega_c^2 + 2 i \beta \omega_c \omega & \kappa^2 \omega_c^2 \\
\kappa^2 \omega_c^2 & \omega^2 - \omega_r^2 + 2 i \alpha \omega_c \omega
\end{array}\right)
\end{equation*}
while Eq.~\eqref{eq:initialHarmonic} (c) becomes 
\begin{equation}
A_\text{out} = -\frac{k_\text{out}^2\omega_c^2}{\omega^2} A_1. \label{eq:aout}
\end{equation}
Solving for $A_1$, $\textbf{A} = \pmb{\Omega}^{-1} \textbf{f}$, the transmission is determined to be

\begin{equation}
|S_\text{21}|^2 = \frac{E_\text{out}}{E_\text{in}} = \eta \frac{\omega_c^8}{\omega^4} \frac{|\omega^2 - \omega_r^2 + 2 i \alpha \omega \omega_c|^2}{|\det\left(\pmb{\Omega}\right)|^2}. \label{eq:oscT}
\end{equation}
Here the amplitude $\eta = (m_\text{out}/m_\text{in}) k_\text{out}^4$ depends on the output coupling $k_\text{out}$ and also the input/output impedance matching determined by $m_\text{out}/m_\text{in}$.  The determinant is given by

\begin{equation}
\det\left(\pmb{\Omega}\right) = \left(\omega^2 - \omega_c^2 + 2 i \beta \omega_c\omega\right)\left(\omega^2 - \omega_r^2 + 2 i \alpha \omega_c \omega\right) - \kappa^4 \omega_c^4. \label{eq:oscDet}
\end{equation}

From Eqs.~\eqref{eq:oscT} and \eqref{eq:oscDet} we can explicitly see the role of the output absorber: it is needed to model the output port and calculate the transmission, since the amplitude of $S_{21}$ in Eq.~\eqref{eq:oscT} is proportional to $k_\text{out}$, however the output plays no role in the dispersion which is determined by the roots of the determinant in Eq.~\eqref{eq:oscDet}.  These facts also illustrate the simplicity of generalizing the model to multiple cavity/multiple spin wave modes: the dispersion and line width will be determined solely from the coupling between any cavity and spin wave modes while the form of the output energy needed to determine the microwave transmission will remain the same as Eq.~\eqref{eq:aout}.

For comparison to the phase correlation and microscopic models discussed later, it is useful to simplify Eqs.~\eqref{eq:oscT} and \eqref{eq:oscDet} by expanding near the coupling point $\omega = \omega_c = \omega_r$.  Near this point $\omega^2 - \omega_r^2 \sim \left(\omega - \omega_r\right) 2 \omega_c$ and therefore Eqs.~\eqref{eq:oscT} and \eqref{eq:oscDet} reduce to
\begin{gather}
|S_{21}|^2 = \eta \frac{\omega_c}{\omega^4} \frac{|\omega - \omega_r + i \alpha \omega |^2}{|\det\left(\pmb{\Omega}\right)|^2} \label{eq:oscTSimp} \intertext{and}
\det\left(\pmb{\Omega}\right) = \left(\omega^2 - \omega_c^2 + 2 i \beta \omega_c \omega\right)\left(\omega-\omega_r + i \alpha \omega\right) - \frac{1}{2}\kappa^4 \omega_c^3. \label{eq:oscDetSimp}
\end{gather}
Comparing to Eq.~\eqref{eq:phaseT} for the phase correlation model we see that the key structure is exactly the same, with differences only in the coupling term which would be expected since the coupling mechanism is implemented differently in the two models.

To verify the agreement between our oscillator model and the experimental CMP features we plot the full result for $|S_{21} \left(\omega, H\right)|^2$ given by Eq.~\eqref{eq:oscDet} in Fig.~\ref{oscFig} (b) using the experimentally determined parameters $\alpha = 0.8 \times 10^{-4}$, $\beta = 3 \times 10^{-4}$ and $\omega_c/2\pi = 10.556$ GHz.  We also use a coupling strength of $\kappa = 0.077$ which is determined experimentally from the value of $\omega_\text{gap}$ (see Sec. \ref{sec:simposc}).  The value of the amplitude $\eta = 2.3 \times 10^{-10}$ determined by the impedance matching is treated as a fitting parameter.  The accuracy of the dispersion and line width determined from Eq.~\eqref{eq:oscT} can immediately be seen by comparing the calculation in Fig.~\ref{oscFig} (b) to the experimental data in Fig.~\ref{dataFig} (b) (more will be said about the dispersion and line width in Sec. \ref{sec:dis}) confirming the accuracy of the harmonic coupling model.  Using the simplified results of Eq.~\eqref{eq:oscTSimp} yields similar agreement.

Eq.~\eqref{eq:oscT} can also be used to calculate the line cuts at fixed field, $|S_{21} \left(\omega\right)|^2$ shown in Fig.~\ref{oscFig} (c) and at fixed frequency, $|S_{21} \left(H\right)|^2$ shown in Fig.~\ref{oscFig} (d).  Comparing the line shapes to the corresponding panels in Fig.~\ref{dataFig} the excellent agreement is evident.  In particular all four of the key line shape features are reproduced by our harmonic coupling model.  In Fig.~\ref{oscFig} (c) we see that i) the fixed $H$ cuts are symmetric in $\omega$ and ii) the amplitude is in good agreement with the experimental results, decreasing as expected above and below the coupling point for the upper and lower branches respectively.  In Fig.~\ref{oscFig} (d) we can see that iii) at $\omega_c$ $|S_{21}\left(H\right)|^2$ has a symmetric dip however above and below $\omega_c$ the line shape has an asymmetry and iv) the polarity of the line shape changes as we pass through the uncoupled cavity mode, in good agreement with the experimental signatures.  Therefore by modelling the FMR/cavity coupling as a set of classical coupled harmonic oscillators we are able to reliably capture the key experimental line shape features of the CMP.
\section{Model II: Dynamic Phase Correlation} \label{sec:phase}

Although the key line shape features of the CMP are captured by a model of harmonic oscillators, this general description does not specify the physical origin of the coupling mechanism.  However recently this question has been answered and the CMP has been shown to result from electrodynamic coupling between cavity and FMR resulting from Faraday's and Amp\`ere's laws \cite{Bai2015, Bai2015b}.  Still, only the CMP dispersion and line width evolution have been investigated in this context.  Here we extend this model to explain the full microwave spectra $S_{21}\left(\omega, H\right)$. 

\begin{figure}[!ht]
\centering
\includegraphics[width=8.5cm]{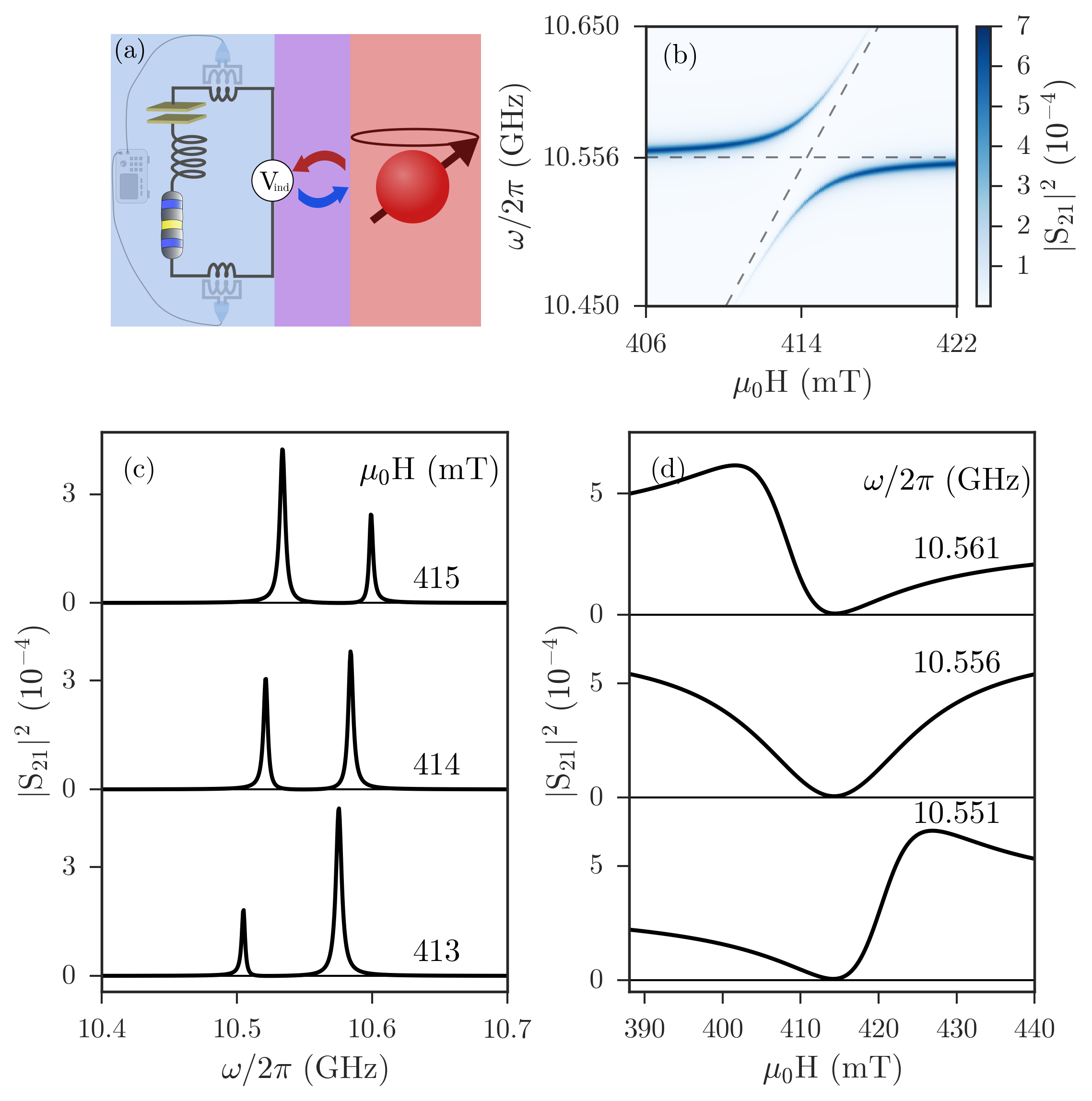}
\caption{Results of the phase correlation model. (a) Schematic illustration of the RLC circuit including the voltage induced by the FMR precession via Faraday's law.  The RLC circuit which models the resonant properties of the microwave cavity is shaded in blue while the FMR precession is shaded in red.  The purple region illustrates the electrodynamic coupling between the two systems. (b) The full $\omega-H$ dispersion calculated according to Eq.~\eqref{eq:phaseT} using the experimental parameters for $\alpha, \beta, \omega_c$ and $\omega_r$ with the coupling $K$ determined from the experimentally measured dispersion gap and an amplitude determined by fitting the experimental data.  (c) Fixed field and (d) frequency cuts made above, at and below the coupling point $\omega_r = \omega_c$ calculated according to Eq.~\eqref{eq:phaseT}.}
\label{phaseFig}
\end{figure}

To build a model centred on the \textit{physical principle} of dynamic phase correlation we use the \textit{technical tool} of microwave circuit theory.  S-parameters are conventionally described using a microwave circuit theory approach \cite{PozarBook} based on the microwave frequency voltages in an RLC circuit like the one shaded in blue in Fig.~\ref{phaseFig} (a).  The complex impedance of such an RLC circuit is 

\begin{equation}
Z_c  = \frac{-i L}{\omega}\left(\omega^2 - \omega_c^2 + 2 i \beta \omega_c \omega\right) \label{eq:circuitImp}
\end{equation}
where $\omega_c^2 = 1/\sqrt{LC}$, $\beta = \Delta\omega/\omega_c = (R/2)(\sqrt{C/L})$ and $R, L, C$ are the resistance, inductance and capacitance of the circuit respectively.  

When a resonant ferromagnetic material is present inside the microwave cavity there is an additional voltage induced due to Faraday's law, $V_x = K_c L dm_y/dt$ and $V_y = - K_c L dm_x/dt$, which can therefore be added into the RLC circuit, at any location, as an additional voltage source.  Moving to a rotational frame, the induced voltage may be written as $V_\text{ind} = -K_cL \omega m^+$ where the solution to the Landau-Lifshitz-Gilbert (LLG) equation is the elliptical magnetization $m^+ = C m_x + i m_y$ with $C = \omega_0/\omega_r$ and $\omega_0 = \gamma H$.  Due to the spherical symmetry of our YIG sample the magnetization is nearly circularly polarized, $C \sim 1$, with any small deviations the result of shape anisotropy, which in our sample is small, $H_A \ll H$. 

On the other hand due to Amp\`eres law the rf magnetic field which drives the magnetization precession may be related to the microwave current in the circuit as $h_x(t) = K_m j_y(t)$ and $h_y(t) = -K_m j_x(t)$ which means that $j^+ = K_m h^+$ with the same definition for the elliptical fields previously used.  Therefore from the LLG equation the magnetization is $m^+ = i \omega_m K_m j^+/\left(\omega- \omega_r + i \alpha \omega\right)$ where $\omega_m = \gamma M_0$.  In our experiment the saturation magnetization is $\mu_0 M_0 = 176$ mT.  Using the magnetization from the LLG equation and the harmonic time dependence $e^{-i\omega t}$, the induced voltage due to the FMR coupling is $V_\text{ind} = \frac{-i \omega_m K^2 L \omega}{\omega- \omega_r + i \alpha \omega} j^+$ and therefore the additional impedance of the RLC circuit due to coupling is
\begin{equation}
Z_m = \frac{-i \omega_m K^2 L \omega}{\omega- \omega_r + i \alpha \omega}
\end{equation}
making the total impedance $Z = Z_c - Z_m$.  Here $K^2 = K_c K_m$ is the total coupling of the CMP system which is related to the coupling in the oscillator model as $K = \kappa^2 \sqrt{\omega_c/2\omega_m}$ (see the discussion of $\omega_\text{gap}$ in Secs. \ref{sec:simposc} and \ref{sec:simpphase}).  In our experiment this means that $K \sim \kappa^2$.  

Using standard microwave network analysis \cite{PozarBook} for our RLC circuit with the additional FMR induced voltage, we find the transmission of the YIG/cavity system to be
\begin{equation}
S_{21} = \frac{2i \omega_c \omega \tilde{\beta} \overline{S_{21}} \left(\omega - \omega_r + i \alpha \omega\right)}{\left(\omega^2 - \omega_c^2 + 2 i \tilde{\beta}\omega \omega_c\right) \left(\omega- \omega_r + i \alpha \omega\right)- \omega^2 \omega_m K^2}. \label{eq:phaseT}
\end{equation}
Here $\tilde{\beta} = \beta + \beta_1 + \beta_2$ is the total loss rate of the cavity which includes the intrinsic losses $\beta$, as well as the loss rates at the input and output ports, $\beta_1$ and $\beta_2$ respectively.  $\overline{S}_{21} = 2 \sqrt{\beta_1 \beta_2}/\tilde{\beta}$ is the maximum amplitude of the uncoupled transmission, $\overline{S}_{21} = S_{21}\left(\omega = \omega_c, K = 0\right)$.  The denominator of Eq.~\eqref{eq:phaseT} is the same as the eigenvalue equation in Ref.~\onlinecite{Bai2015} and its roots determine the dispersion and line width of the CMP.

Fig.~\ref{phaseFig} (b) shows the calculation of $|S_{21} \left(\omega, H\right)|^2$ based on Eq.~\eqref{eq:phaseT} using the experimentally determined parameters for $\alpha, \beta, \omega_c$ and $\omega_m$.  In this calculation we have also used the coupling strength $K = 0.006$ measured from the dispersion gap as described in Sec. \ref{sec:simpphase} and determined $\tilde{\beta} \overline{S}_{21} = 7.6 \times 10^{-6}$ from a fit to the experimental data.  We find excellent agreement when comparing the dispersion and line width from Fig.~\ref{phaseFig} (b) to the experimental data in Fig.~\ref{dataFig} (b).  This is expected since, as mentioned, the roots of the denominator in Eq.~\eqref{eq:phaseT} are equivalent to the CMP eigenvalue solutions which were previously shown to describe the dispersion\cite{Bai2015}.

To check the line shapes within this model we use Eq.~\eqref{eq:phaseT} to calculate $|S_{21} \left(\omega\right)|^2$ at fixed $H$ and $|S_{21} \left(H\right)|^2$ at fixed $\omega$ as shown in Figs. \ref{phaseFig} (c) and (d) respectively.  When these figures are compared to Figs. \ref{dataFig} (c) and (d) all of the key line shape features are seen to agree.  In particular from Fig.~\ref{phaseFig} (c) we see that i) $|S_{21} \left(\omega\right)|^2$ is symmetric for all fixed $H$ and ii) the amplitude is decreasing as expected above and below the coupling point for the upper and lower branches respectively.  From Fig.~\ref{phaseFig} (d) we can see that iii) $|S_{21}\left(H\right)|^2$ is only symmetric at $\omega_c$ and has a dip, however immediately above and below $\omega_c$ the line shape is asymmetric and iv) the polarity of the asymmetry changes as we pass through $\omega_c$, in agreement with the experimental features.  This excellent agreement between the experimental line shape and the dynamic phase correlation model further confirms that the origin of the CMP is electrodynamic coupling and provides a solid physical model for the analysis and fitting of the transmission spectra in strongly coupled magnon-photon systems.

\section{Model III: Microscopic Theory} \label{sec:greens}
The previous two models were based on a macroscopic, classical description. A microscopic theory, on the other hand, necessarily has to be a quantum theory of the spins, the electromagnetic field, and the spin-photon coupling. In the following we want to develop such a theory and discuss the approximations which lead to an effective theory of two coupled quantum harmonic oscillators.   

\subsection{Hamiltonian}
For the experimentally relevant energy scales, the physical properties
of YIG can be understood in terms of an effective spin-s Heisenberg
ferromagnet on a cubic lattice. The effective Hamiltonian contains
both exchange as well as dipole-dipole interactions and can be written
as\cite{Kreisel2009}
\begin{equation}
\label{H_YIG}
H_s=-\frac{1}{2}\sum_{ij}\sum_{\alpha\beta} \left[J_{ij}\delta^{\alpha\beta}+D_{ij}^{\alpha\beta}\right] S_i^\alpha S_j^\beta -g\mu_B B_z\sum_j S^z_j.
\end{equation}
Here $J_{ij}= J$ for nearest neighbors while $J_{ij}\approx 0$
otherwise. $D_{ij}^{\alpha\beta}$ is the dipolar tensor. We have also
included a constant external magnetic field $B_z$ pointing along the
$z$-direction. The Hamiltonian for the microwave field inside the
cavity is simply given by
\begin{equation}
\label{photon}
H_{\textrm{ph}} =\hbar \sum_q \omega_q (a_q^\dagger a_q +1/2).
\end{equation}
Finally, we have to consider the coupling between the rf field and the
spins
\begin{equation}
\label{int}
H_{\textrm{int}} =g\mu_B \sum_j \overline{\vec{B}}_j\cdot\vec{S}_j.
\end{equation}
By $\overline{\vec{B}}\equiv(B_x,B_y,0)$ we denote the circular
polarized rf field which can be quantized using standard
techniques\cite{WallsBook} leading to
\begin{align*}
\overline{\vec{B}} = \frac{1}{c} \sum_q &\sqrt{\frac{\hbar \omega_q}{4 \epsilon_0 V}} \left[\left(a_q e^{i {\bf q}\cdot {\bf r}} + a_q^\dagger e^{-i {\bf q}\cdot {\bf r}}\right) \widehat{\bf x} \right.\\
&+ \left.i \left(a_q e^{i {\bf q}\cdot {\bf r}} - a_q^\dagger e^{-i {\bf q}\cdot {\bf r}}\right) \widehat{\bf y}\right].
\end{align*}
Here $V$ is the cavity volume and $\epsilon_0$ is the vacuum
permittivity. Note that we have not attempted to expand the field in a
basis which takes the proper boundary conditions due to the shape of
the cavity into account but rather assumed that we can approximate the
field in a plane-wave basis near the position of the sample. Using the
ladder operators $S_j^\pm =S^x_j\pm i S^y_j$ the interaction between
the spin and the photon system, Eq.~\eqref{int}, can then be written
as
\begin{equation}
\label{int2}
H_{\textrm{int}} =\frac{g\mu_B}{c} \sum_{j,q} \sqrt{\frac{\hbar \omega_q}{4 \epsilon_0 V}} \left(a_qS^+_j e^{i {\bf q}\cdot {\bf r_j}} + a_q^\dagger S^-_j e^{-i {\bf q}\cdot {\bf r_j}}\right).
\end{equation}

The microscopic quantum Hamiltonian of the spin-cavity system is
therefore given by $H=H_s+H_{\textrm{ph}}+H_{\textrm{int}}$. 

In good approximation only the lowest magnon band of the Hamiltonian
\eqref{H_YIG} is important. We can therefore apply linear spin-wave
theory to simplify the problem. The spin operators are expressed in
terms of bosonic operators $b^{(\dagger)}$ in the following way
\begin{equation}
\label{HP}
S^z_j= b_j^\dagger b_j -s, \quad S^+_j=\sqrt{2s}b^\dagger,\quad S^-_j=\sqrt{2s}b_j.
\end{equation}
Using a Fourier transform $b_j=\frac{1}{\sqrt{N_s}}\sum_{k}
b_{k}\e^{ikr}$ the Hamiltonian of the spin-cavity system then reads
\begin{eqnarray}
\label{Ham_spin_cavity}
H&=&\hbar \sum_k\omega^s_k (b_k^\dagger b_k +1/2) +\hbar \sum_q \omega_q (a_q^\dagger a_q +1/2) \nonumber \\
&+& \hbar \sqrt{N_s}\alpha_c \sum_k (a_k b_k^\dagger +a_k^\dagger b_k) 
\end{eqnarray}
with a coupling constant
\begin{equation}
\label{coupling}
\alpha_c=\frac{g\mu_B}{\hbar c}\sqrt{\frac{s\hbar\omega_q}{2\epsilon_0 V}}.
\end{equation}
Here $\omega_k^s$ is the magnon dispersion. In linear spin-wave theory
we therefore obtain a system of coupled harmonic oscillator modes.

\begin{figure}[!ht]
\centering
\includegraphics[width=8.5cm]{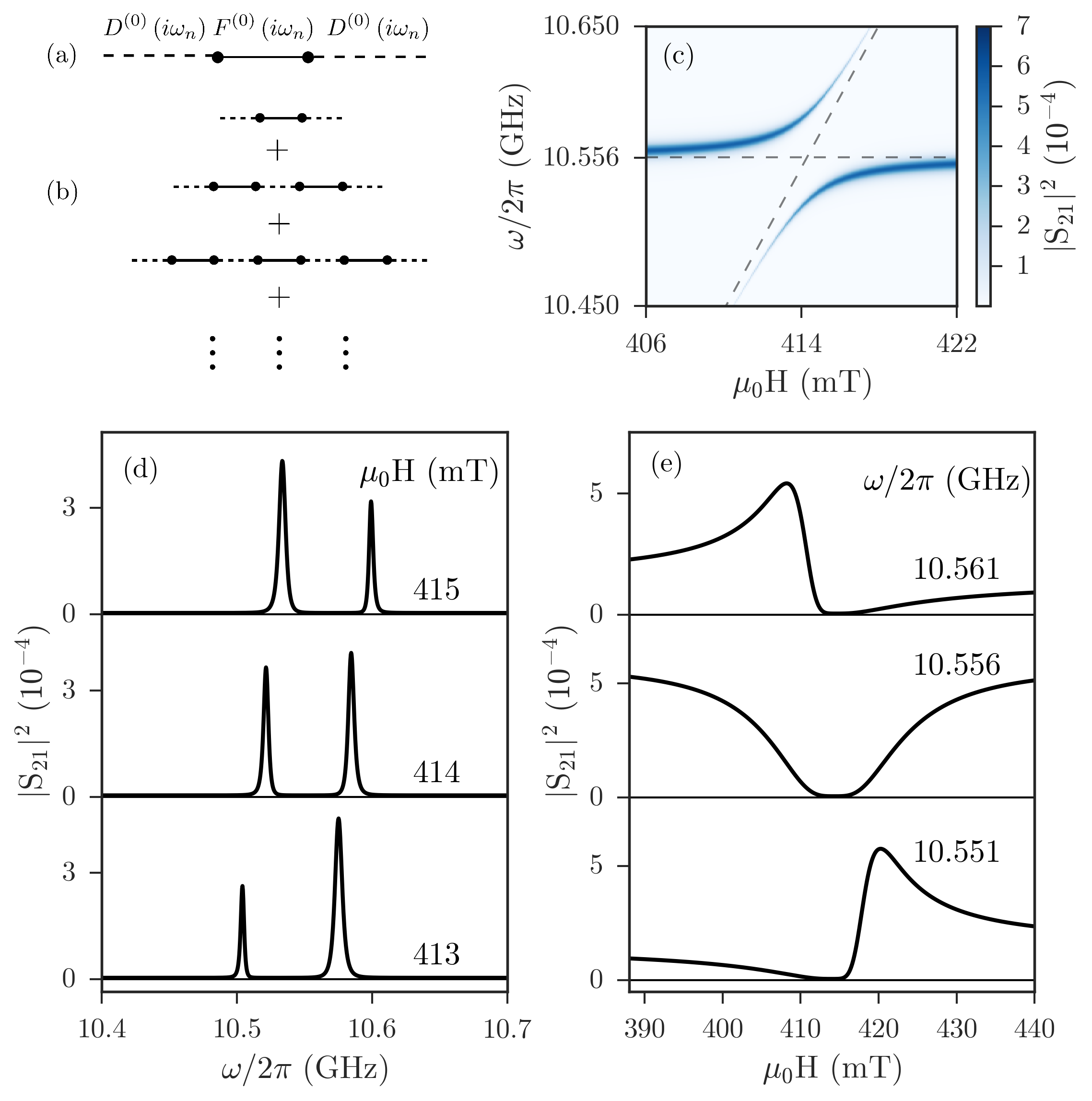}
\caption{Results of the Green's function formalism. (a) Scattering of a photon off an FMR magnon. (b) The geometric series of diagrams can be summed to obtain the cavity photon Green's function \eqref{fullGF}.  (d) The full $\omega-H$ dispersion calculated according to Eq.~\eqref{eq:greensT} using the experimental parameters for $\alpha, \beta, \omega_c$ and $\omega_r$ with the coupling $\alpha_c$ determined from the experimentally measured dispersion gap and an amplitude determined by fitting the experimental data.  (d) Fixed field and (e) frequency cuts made above, at and below the coupling point $\omega_r = \omega_c$ calculated according to Eq.~\eqref{eq:greensT}.}
\label{greensFig}
\end{figure}

A further drastic simplification can be achieved if we consider only a
single cavity mode $\omega_c$ and only the FMR mode $\omega^s_{k\to
0}=\omega_r$. In this case, we end up with a Dicke model (or
Tavis-Cummings model)
\begin{equation}
\label{Dicke}
H =  \hbar \omega_r b^\dagger b+ \hbar \omega_c a^\dagger a + \hbar \sqrt{N_s}\alpha_c \left(a b^\dagger + a^\dagger b\right)
\end{equation}
which is an $N_s$-spin version of the Jaynes-Cummings
model.\cite{Garraway2011} The model describes two coupled quantum
harmonic oscillators: one associated with the Dicke spin and the other
with the electromagnetic field. Note that the excitation number
$M=a^\dagger a+b^\dagger b$ is conserved, $[H,M]=0$. The Hamiltonian
\eqref{Dicke} can be diagonalized by a Bogoliubov transformation
\begin{equation}
\label{Bogoliubov}
a=-\sin\beta c_1 + \cos\beta c_2,\quad b=\cos\beta c_1+\sin\beta c_2
\end{equation}
where $c_{1,2}$ are two new bosonic operators. The angle $\beta$ which
diagonalizes the problem is given by $\tan(2\beta)=
2\sqrt{N_s}\alpha_c/(\omega_c-\omega_r)$. The eigenspectrum
$\varepsilon_{n,j}=\hbar
\omega_{n,j}$ of the coupled oscillators is given by

\begin{equation}
\label{eigen}
\omega_{n,j} = \frac{\omega_c+\omega_r}{2}n+j\sqrt{(\omega_r-\omega_c)^2+4N_s\alpha_c^2}
\end{equation}
with $n=0,1,\cdots,N_s$ and $j\in[-\frac{n}{2},\frac{n}{2}]$. The lowest
polariton modes ($n=1$) are, in particular, given by 
\begin{equation}
\label{eigen2}
\omega_\pm = \frac{1}{2}\left[\omega_c+\omega_r\pm\sqrt{(\omega_r-\omega_c)^2+4N_s\alpha_c^2}\right].
\end{equation}
\subsection{Transmission}
We are interested in calculating the microwave transmission through
the cavity. This can be understood as a scattering problem and we
require the matrix element $S_{21}$ of the scattering matrix $S$. For
an incoming state $|i\rangle$ and and outgoing state $|f\rangle$ we
can relate the S matrix to the $\mathcal{T}$ matrix by
\begin{equation}
\label{T-matrix}
\langle f|S|i\rangle =\delta_{fi}+2\pi i\delta(E_f-E_i)\langle f|\mathcal{T}|i\rangle.
\end{equation}
If we couple our cavity to a microwave field we have to consider an
open quantum system where the microwave photons form a 'bath' coupling
for the cavity photons. By integrating out the bath photons we obtain
$\mathcal{T}(\omega)\sim \lambda^2 D^{\textrm{ret}}(\omega)$ where
$\lambda$ is the coupling constant between microwave bath and cavity
photons (assumed to be frequency independent) and
$D^{\textrm{ret}}(\omega)$ is the retarded Green's function of the
{\it cavity photons}. I.e., the transmission can be calculated from
the properties of the spin-cavity system alone.\cite{DattaBook} For
the transmission amplitude, Eq.~\eqref{T-matrix} implies
$t(\omega)=2\pi i\rho(\omega)\mathcal{T}(\omega)$ where $\rho(\omega)$
is the density of states of the external microwave
photons.\cite{Chirla2016} If we assume $\rho(\omega)$ to be
constant near the resonance frequency and ignore the asymmetric
contribution from the real part of the Green's function we obtain
\begin{equation}
\label{transmission}
|S_{21}(\omega)|^2\propto -\Im D^{\textrm{ret}}(\omega).
\end{equation}
The measured transmission spectra is therefore approximately given by
the spectral function of the cavity photons.

\subsection{Photon Green's function}
For the decoupled system---Eq.~\eqref{Dicke} with $\alpha_c=0$---the
Matsubara Green's function for the photons is given by
\begin{eqnarray}
\label{GF1}
D^{(0)}(i\omega_n)&=&\langle (a+a^\dagger)(a+a^\dagger)\rangle(i\omega_n)  \\
&=&\frac{1}{i\omega_n-\omega_c+i\Gamma_c}-\frac{1}{i\omega_n+\omega_c+i\Gamma_c} \nonumber
\end{eqnarray}
and for the magnons by 
\begin{eqnarray}
\label{GF2}
F^{(0)}(i\omega_n)&=&\langle (b+b^\dagger)(b+b^\dagger)\rangle(i\omega_n) \\
&=&\frac{1}{i\omega_n-\omega_r+i\Gamma_r}-\frac{1}{i\omega_n+\omega_r+i\Gamma_r}. \nonumber
\end{eqnarray}
Here we have introduced two damping rates: $\Gamma_c$ related to
cavity losses and $\Gamma_r$ due to the intrinsic Gilbert damping of
the FMR resonance. We will not try to calculate $\Gamma_{c,r}$ but
rather use them as fitting parameters. The Green's function of the
cavity photons for the coupled system, Eq.~\eqref{Dicke}, can be
obtained straightforwardly by expressing the photon creation and
annihilation operators in terms of the diagonal basis
$c_{1,2}^{(\dagger)}$ using the Bogoliubov transform
\eqref{Bogoliubov}. An alternative way to obtain the full Green's
function $D(i\omega_n)$ is to calculate the photon self energy from
Dyson's equation. For the Dicke model there is only a single
photon-magnon diagram so that the scattering series can be calculated
exactly, see Fig.~\ref{greensFig}(a,b). The result for the retarded
Green's function, obtained after analytic continuation, is given by
\begin{eqnarray}
\label{fullGF}
D^{\textrm{ret}}(\omega)&=&\left(\omega-\omega_c+i\Gamma_c-\frac{N\alpha_c^2}{\omega-\omega_r+i\Gamma_r}\right)^{-1} \nonumber \\
&-&\left(\omega+\omega_c+i\Gamma_c-\frac{N\alpha_c^2}{\omega+\omega_r+i\Gamma_r}\right)^{-1}.
\end{eqnarray}
The final result for the transmission spectra therefore is
\begin{widetext}
\begin{eqnarray}
\label{eq:greensT}
&& |S_{21}|^2 \propto -\text{Im} D^{\textrm{ret}}(\omega) \\
&=& \frac{ \Gamma_c + \frac{ N\alpha_c^2 \Gamma_r}{\left(\omega - \omega_r\right)^2+ \Gamma_r^2}}{\left(\omega - \omega_c -\frac{N\alpha_c^2(\omega - \omega_r)}{\left(\omega - \omega_r \right)^2 + \Gamma_r^2}\right)^2 + \left(\Gamma_c + \frac{N\alpha_c^2\Gamma_r}{\left(\omega - \omega_r\right)^2 + \Gamma_r^2}\right)^2}
-\frac{ \Gamma_c + \frac{ N\alpha_c^2 \Gamma_r}{\left(\omega + \omega_r\right)^2+ \Gamma_r^2}}{\left(\omega + \omega_c -\frac{N\alpha_c^2(\omega + \omega_r)}{\left(\omega + \omega_r \right)^2 + \Gamma_r^2}\right)^2 + \left(\Gamma_c + \frac{N\alpha_c^2\Gamma_r}{\left(\omega + \omega_r\right)^2 + \Gamma_r^2}\right)^2}. \nonumber
\end{eqnarray}
\end{widetext}
We define the proportionality constant to be
$\overline{S}_{21}^{~2}\Gamma_c$ where $\overline{S}_{21} =
S_{21}\left(\alpha_c = 0, \omega = \omega_c\right)$.

Fig.~\ref{greensFig} (c) shows the calculation of $|S_{21} \left(\omega, H\right)|^2$ based on Eq.~\eqref{eq:greensT} using the experimentally determined parameters for $\alpha, \beta$ and $\omega_c$ taking $\Gamma_r = \omega_c \alpha$ and $\Gamma_c = \omega_c \beta$.  In this calculation we have also used the amplitude $\overline{S}_{21} = 0.048$, determined from a fit to the experimental data, and the coupling strength $\sqrt{N_s}\alpha_c = 32$ MHz measured from the dispersion gap as described in Sec. \ref{sec:simpgreens}.  We note that the value of 32 MHz agrees very well with the value of 37 MHz calculated directly from Eq.~\eqref{coupling}.  We find excellent agreement when comparing the dispersion and line width from Fig.~\ref{greensFig} (c) to the experimental data in Fig.~\ref{dataFig} (b).  Figs. \ref{greensFig} (d) and (e) show $|S_{21} \left(\omega\right)|^2$ at fixed $H$ and $|S_{21} \left(H\right)|^2$ at fixed $\omega$ respectively.  When these figures are compared to Figs. \ref{dataFig} (c) and (d) all of the key line shape features are seen to agree.  In particular from Fig.~\ref{greensFig} (d) we see that i) $|S_{21} \left(\omega\right)|^2$ is symmetric for all fixed $H$ and ii) the amplitude is decreasing as expected above and below the coupling point for the upper and lower branches respectively.  From Fig.~\ref{greensFig} (e) we can see that iii) $|S_{21}\left(H\right)|^2$ is only symmetric at $\omega_c$ and has a dip, however immediately above and below $\omega_c$ the line shape is asymmetric and iv) the polarity of the asymmetry changes as we pass through $\omega_c$, in agreement with the experimental features. 
\section{Line Shape Simplification}\label{sec:simp}

While the transmission functions in Eq.~\eqref{eq:oscT}, Eq.~\eqref{eq:phaseT} and Eq.~\eqref{eq:greensT} accurately describe the CMP line shape, these equations are not simple Lorentz and asymmetric functions.  This is expected, since even in the simple RLC case without the FMR induced voltage, the S parameters are only Lorentzian when expanded near the resonance frequency \cite{Petersan1998}.  In order to provide a further simplified equation for the fitting of experimental data we now expand these equations near the coupled resonances.

\subsection{Model I: Harmonic Coupling} \label{sec:simposc}
To simplify the line shape we first need an analytic approximation for the eigenmodes.  Since the losses are very small, $\alpha, \beta \ll 1$, the damping has a negligible effect on the dispersion and can be safely ignored so that, denoting the upper and lower branches by $\omega_+$ and $\omega_-$ respectively, the roots of Eq.~\eqref{eq:oscDetSimp} near the coupling point $\omega = \omega_c$ are
\begin{equation}
\omega_\pm = \frac{1}{2} \left[ \left(\omega_r + \omega_c\right) \pm \sqrt{\left(\omega_c-\omega_r\right)^2 + \kappa^4 \omega_c^2}\right]. \label{eq:oscEigen}
\end{equation}
This approximation is in excellent agreement with the numerical results shown in Fig.~\ref{phaseDisFig} (a).  An important application of Eq.~\eqref{eq:oscEigen} is to determine the dispersion gap.  Setting $\omega_r = \omega_c$ we find $\omega_\text{gap} = \omega_+ - \omega_- = \kappa^2 \omega_c$.  This allows the coupling strength $\kappa$ to be easily determined directly from the experimental data, without any fitting.

Expanding Eq.~\eqref{eq:oscTSimp} to second order near the CMP eigenmodes,
\begin{equation}
|S_{21}^\pm\left(\omega\right)|^2 = \frac{N_0^\pm + N_1^\pm \left(\omega - \omega_\pm\right) + N_2^\pm \left(\omega - \omega_\pm\right)}{D_0^\pm + D_1^\pm \left(\omega - \omega_\pm\right) + D_2^\pm \left(\omega - \omega_\pm\right)} \label{eq:s21Expand}
\end{equation}
where $S_{21}^+$ and $S_{21}^-$ are the expansions near the upper and lower branches respectively.  For $|\omega_r - \omega_\pm| < \beta\omega_c$ we find that $D_1^\pm \approx 0$ and $N_0^\pm >> N_1^\pm, N_2^\pm$ which means that the transmission near either resonances, $\omega_e = \omega_\pm$ can be written as
\begin{equation}
|S_{21}\left(\omega\right)|^2 = A \frac{\Delta \omega}{\Delta \omega^2 + \left(\omega - \omega_e\right)^2}. \label{eq:Lorentz}
\end{equation}
This shows that, within the oscillator model, at fixed field the line shape of the CMP near resonance, $|S_{21}\left(\omega\right)|^2$, is simply Lorentzian.  

The condition $|\omega_r - \omega_\pm| < \beta\omega_c$ ensures that the CMP eigenmode is sufficiently separated from the uncoupled FMR frequency.  This is required to avoid any unwanted distortions from the expected Lorentzian line shape.  Eq.~\eqref{eq:oscT} shows that the full expression for $|S_{21}\left(\omega, H\right)|^2$ has an antiresonance (minimum) when $\omega = \omega_r$ which can cause a distortion in the line shape as $\omega_\pm$ approaches $\omega_r$.  This effect is accentuated by the fact that at this point the amplitude is already greatly diminished.  In our case this constraint corresponds to $\mu_0 H < 426$ mT for the upper branch and $\mu_0 H > 402$ mT for the lower branch.  In these regimes the transmission amplitude is too small to perform reliable fits, being several orders of magnitude smaller than the main resonance, and therefore in practice the distortion due to antiresonance does not limit the applicability of Eq.~\eqref{eq:Lorentz} and the use of Lorentzian fits.

To perform a similar analysis for $|S_{21}\left(H\right)|^2$ we simply solve for the $\omega_r$ roots of the determinant in Eq.~\eqref{eq:oscDet} and expand $|S_{21}\left(\omega, H\right)|^2$ to second order near this point.  The form of the expansion is the same as the one just given in Eq.~\eqref{eq:s21Expand}.  Again in this case $D_0 \approx 0$, however $N_1$ and $N_2$ are no longer negligible.  This means that $|S_{21}\left(H\right)|^2$ has the form
\begin{equation}
|S_{21}\left(H\right)|^2 =  A\frac{\left(q \Delta \omega + \omega - \omega_e\right)^2}{\left(\omega-\omega_e\right)^2 + \Delta \omega^2}. \label{eq:assym}
\end{equation}
The simplification can be used in order to fit the $|S_{21}\left(H\right)|^2$ cuts, taking $A, q, \omega$ and $\omega_e$ as fitting parameters.  The parameter $q$ is a function of $\alpha, \beta, \omega_c$ and $\omega$ and controls the degree of asymmetry.  It is approximately linear over the narrow frequency range where coupling is observed and its most important characteristic is that $q\left(\omega < \omega_c\right) < 0, q\left(\omega = \omega_c\right) = 0$ and $q\left(\omega > \omega_c\right) >0$.  This accounts for the change in symmetry observed in Fig.~\ref{dataFig} (d).   
\subsection{Model II: Dynamic Phase Correlation} \label{sec:simpphase}
The procedure to simplify Eq.~\eqref{eq:phaseT} is the same as that used in the previous subsection.  In this case, near the coupling point $\omega = \omega_c$ and for small coupling and damping, $K^2, \alpha, \beta \ll 1$, we can approximate the eigenvalues by

\begin{equation}
\omega_\pm = \frac{1}{2}\left[\omega_r + \omega_c \pm \sqrt{\left(\omega_c-\omega_r\right)^2 + 2 K^2 \omega_m\omega_c}\right]. \label{eq:phaseDisp}
\end{equation}
Note that small $K^2$ here means small compared to 1, not small compared to $\alpha$ and $\beta$ so this approximation is still valid within the strong coupling regime.  Again this analytic expression for the eigenmodes allows us to determine the dispersion gap, $\omega_\text{gap} = K \sqrt{2 \omega_m \omega_c}$, which enables the electrodynamic coupling strength to to be determined directly from the experimental data without treating it as a fitting parameter.  

Expanding near the eigenmodes the behaviour for the expansion coefficients is the same as that found for the harmonic coupling model, and therefore it is again appropriate to use the Lorentz line shape in Eq.~\eqref{eq:Lorentz} to fit the constant field cuts, $|S_{21}\left(\omega\right)|^2$, and to use the asymmetric line shape in Eq.~\eqref{eq:assym} to fit the constant frequency cuts, $|S_{21}\left(H\right)|^2$.

\subsection{Model III: Microscopic Theory} \label{sec:simpgreens}

Again, to simplify Eq.~\eqref{eq:greensT} we take the same approach. The spectral function of the cavity photons is peaked at the polariton modes $\omega_\pm$, see Eq.~\eqref{eigen2}, for small damping. Eq.~\eqref{eigen2} is exactly the same as Eq.~\eqref{eq:phaseDisp} with $\omega_\text{gap} = 2\sqrt{N_s}\alpha_c$ which means that $\alpha_c = K \sqrt{\frac{\omega_m \omega_c}{2N_s}}$. Expanding near the eigenmodes the behaviour for the expansion coefficients is the same as that found for the previous models, and therefore it is again appropriate to use the Lorentz line shape in Eq.~\eqref{eq:Lorentz} to fit the constant field cuts, $|S_{21}\left(\omega\right)|^2$, and to use the asymmetric line shape in Eq.~\eqref{eq:assym} to fit the constant frequency cuts, $|S_{21}\left(H\right)|^2$.

With analytic forms of the dispersion, the coupling parameters of each model can be related,
\begin{equation*}
\omega_\text{gap} = \frac{g\mu_B}{\hbar c}\sqrt{\frac{2 s N_s \hbar \omega_c}{\epsilon_0 V}} = \kappa^2 \omega_c = K \sqrt{2 \omega_m \omega_c}.
\end{equation*}
In particular this allows us to determine the coupling strengths of the classical oscillator and phase correlation model in terms of the microscopic parameters,
\begin{figure}[!h]
\centering
\includegraphics[width=8.5cm]{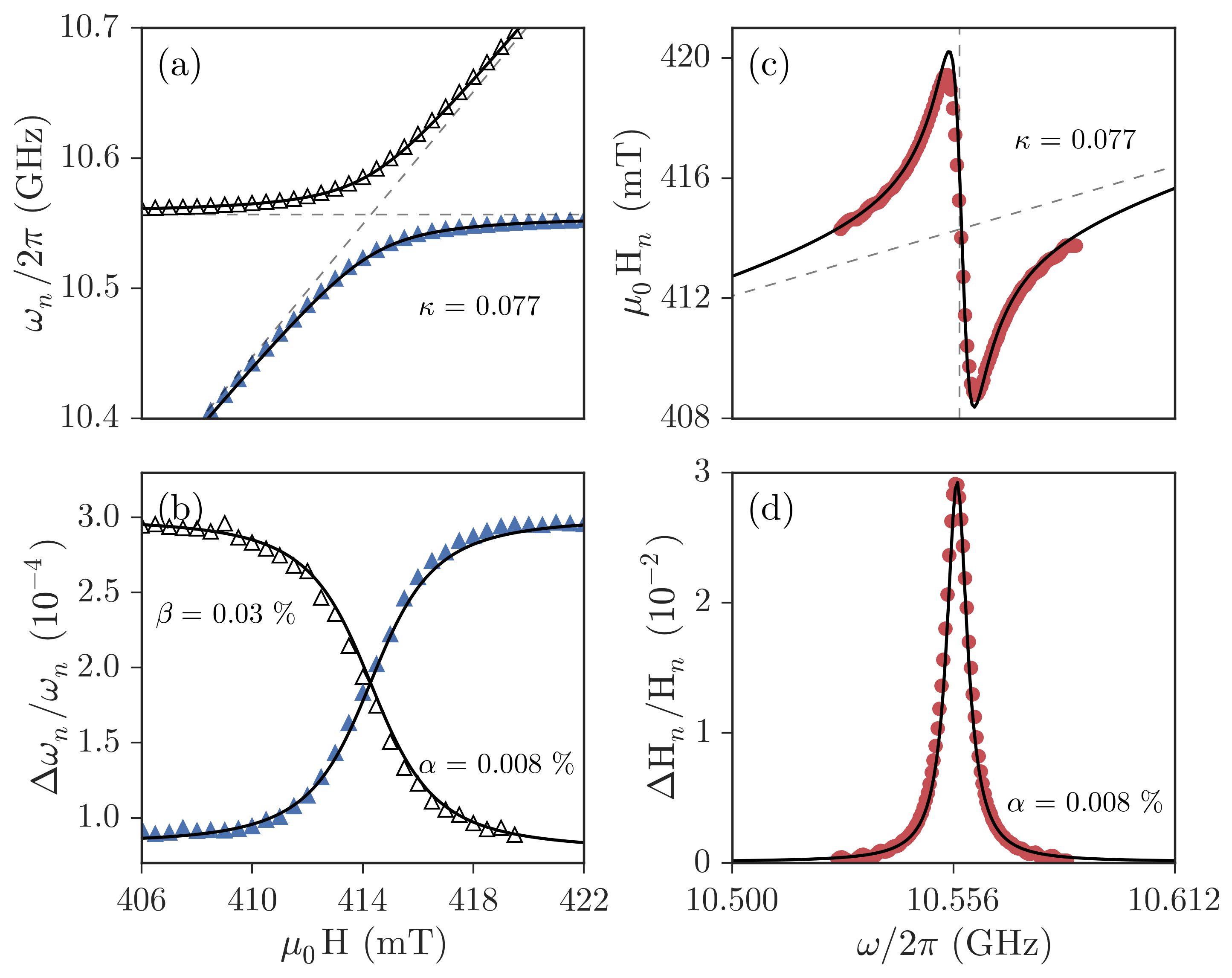}
\caption{Cavity-magnon-polariton dispersion and line width compared to the oscillator model.  (a) The resonance position and (b) the line width (HWHM) determined from $|S_{21}\left(\omega\right)|^2$.  Symbols are fit results according to Eq.~\eqref{eq:Lorentz} while the solid curves are calculated by finding the complex $\tilde{\omega}_n$ roots of Eq.~\eqref{eq:oscDet}.  Horizontal and diagonal dashed lines in (a) indicate the uncoupled cavity and FMR dispersions respectively.  (c) The resonance position and (d) the line width determined from $|S_{21}\left(H\right)|^2$.  Symbols are fit results according to Eq.~\eqref{eq:assym} while the solid curves are calculated by finding the complex $\tilde{H}_n$ root of Eq.~\eqref{eq:oscDet}.  Vertical and diagonal dashed lines in (c) indicate the uncoupled cavity and FMR dispersions respectively.}
\label{oscillatorDisFig}
\end{figure}
\begin{equation*}
\kappa^2 = \frac{g \mu_B}{\hbar c} \sqrt{\frac{2s N_s \hbar}{\epsilon_0 \omega_c V}}, ~~ K = \frac{g \mu_B}{\hbar c}\sqrt{\frac{s N_s \hbar}{\epsilon_0 \omega_m V}}.
\end{equation*}
%
%
\section{Dispersion and line width}\label{sec:dis}
\begin{figure}[!h]
\centering
\includegraphics[width=8.5cm]{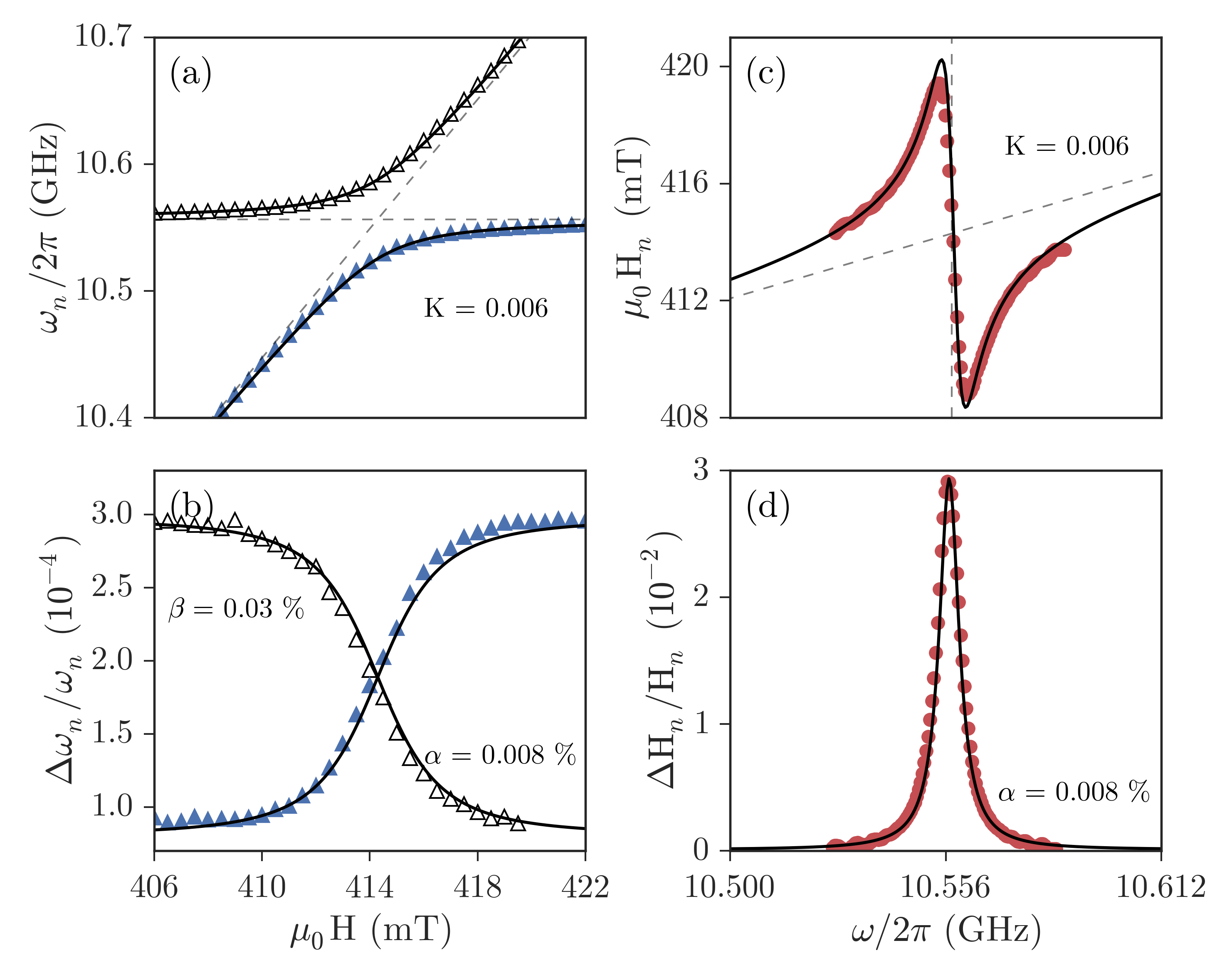}
\caption{Cavity-magnon-polariton dispersion and line width compared to the phase correlation model.  (a) The resonance position and (b) the line width (HWHM) determined from $|S_{21}\left(\omega\right)|^2$.  Symbols are fit results according to Eq.~\eqref{eq:Lorentz} while the solid curves are calculated by finding the complex $\tilde{\omega}_n$ roots of the denominator in Eq.~\eqref{eq:phaseT}.  The horizontal and diagonal dashed lines in (a) indicate the uncoupled cavity and FMR dispersions respectively.  (c) The resonance position and (d) the line width determined from $|S_{21}\left(H\right)|^2$.  Symbols are fit results according to Eq.~\eqref{eq:assym} while the solid curves are calculated by finding the complex $\tilde{H}_n$ root of the denominator in Eq.~\eqref{eq:phaseT}.  The vertical and diagonal dashed lines in (c) indicate the uncoupled cavity and FMR dispersions respectively.  Note that the solid curves shown here for the phase correlation model are exactly the same as those calculated in the microscopic theory using the Green's function formalism.}
\label{phaseDisFig}
\end{figure}

To quantitatively confirm both the applicability of the line shape approximations described in the previous section, and the accuracy of the dispersion and line width in each model, we have fit the full experimental data set for both fixed $\omega$ and fixed $H$.  The symbols in Fig.~\ref{oscillatorDisFig} and Fig.~\ref{phaseDisFig} (a) and (b) show the fitting results for the dispersion and HWHM according to the Lorentz function in Eq.~\eqref{eq:Lorentz}.  The open symbols in both (a) and (b) correspond to the high frequency branch while the solid symbols correspond to the low frequency branch.  The horizontal and vertical dashed lines in (a) are the uncoupled cavity and FMR dispersions respectively.  

The solid curves in Fig.~\ref{oscillatorDisFig} (a) and (b) are solutions for the roots of the determinant in Eq.~\eqref{eq:oscDet} from the oscillator model while the solid curves in Fig.~\ref{phaseDisFig} (a) and (b) are solutions for the roots of the denominator in Eq.~\eqref{eq:phaseT} from the phase correlation model which are identical to the Green's function calculation 
(which is therefore not shown).  In both cases $\omega_n = \text{Re}\left(\tilde{\omega}_n\right)$ with $n = 1, 2$ labelling the two complex solutions $\tilde{\omega}_n$ and $\Delta \omega_n = \text{Im}\left(\tilde{\omega}_n\right)$.  There is excellent agreement between the fit results and the expected dispersion and line width evolution, with the damping of the system evolving between the pure cavity loss rate of $\beta = 0.03 \%$ and the pure YIG loss rate of $\alpha = 0.008 \%$ (the limiting behaviour of these fits is the experimental method used to determine the Gilbert damping).    

As expected we see that the damping of the upper and lower branches is the same, and equal to the average loss rate $\left(\beta + \alpha\right)/2$ at the coupling point $\omega_r = \omega_c$.  

The resonant position and HWHM at fixed frequency are shown in Fig.~\ref{oscillatorDisFig} and Fig.~\ref{phaseDisFig} (c) and (d) determined from the $|S_{21}\left(H\right)|^2$ fits using Eq.~\eqref{eq:assym}.  The vertical and diagonal dashed lines in (c) show the uncoupled cavity and FMR dispersions respectively.  The complex $\tilde{H}_n$ roots of Eq.~\eqref{eq:phaseT} ($n=1$) are shown as solid curves in (c) and (d) with $H_n = \text{Re}(\tilde{H}_n)$ and $\Delta H_n = \text{Im}(\tilde{H}_n)$.  Again (c) and (d) show good agreement between the theoretical curves from both the oscillator model and the circuit theory model with the experimental data.
\section{Conclusions}
We have presented a detailed study of the cavity-magnon-polariton microwave transmission line shape measured by inserting a magnetically ordered material into a microwave cavity.  At fixed field we found that the transmission $|S_{21}\left(\omega\right)|^2$ has symmetric resonances while the fixed frequency line shape $|S_{21}\left(H\right)|^2$ is generally asymmetric and the polarity of this asymmetry can be controlled by the driving microwave frequency.  These line shape observations are important to extract the magnetic characteristics of the CMP ($H_r$ and $\Delta H$) from the transmission spectra and may play a role in the phase coherent control of the CMP.  

To describe our observations we developed three models of the CMP.  First, for the first time we directly compared the transmission spectra to a coupled oscillator model and found that all of the important line shape features are accurately described.  Second, we explored the detailed physical nature of this coupling and found that the unique CMP line shape has a classical electrodynamic origin which can be implemented quantitatively using microwave circuit theory.  Finally we performed a Green's function calculation in a microscopic quantum model. In linear spin-wave theory and taking only the FMR and a single cavity mode into account we again found an accurate description of the CMP line shape.  In addition to the excellent line shape agreement, all three models accurately describe the dispersion and line width evolution of the CMP and allow the coupling strength to be extracted directly from the dispersion gap, meaning that the important coupling parameter can be determined without fitting.  In the low damping limit where strong coupling experiments are performed, each model provides the same simple analytic line shape formula which can be used to quantitatively analyze a wide range of strongly coupled spin-photon systems currently employed within the community.  Due to these advantageous features we believe that the experimental observations and theoretical approaches we have presented provide a useful quantitative tool to describe currently studied strongly coupled spin-photon systems and can be easily extended to multi mode systems or to include higher order interaction terms. By explaining the CMP line shape these models provide a step towards further coherent control of the cavity-magnon-polariton.
\\[10pt]
\section*{ACKNOWLEGEMENTS}
We thank Z. H. Zhang for useful discussions.  M.H. is supported by the NSERC CGSD program.  This work has been funded by NSERC (J. S., C.-M. H), as well as NSFC (No. 11429401), CFI and CMC grants (C.-M. H.). 
\bibliography{mainText.bbl}
\end{document}